# GPU Sample Sort

Nikolaj Leischner* Vitaly Osipov† Peter Sanders‡

**Abstract**

In this paper, we present the design of a sample sort algorithm for manycore GPUs. Despite being one of the most efficient comparison-based sorting algorithms for distributed memory architectures its performance on GPUs was previously unknown. For uniformly distributed keys our sample sort is at least $25\%$ and on average $68\%$ faster than the best comparison-based sorting algorithm, GPU Thrust merge sort, and on average more than $2$ times faster than GPU quicksort. Moreover, for $64$-bit integer keys it is at least $63\%$ and on average $2$ times faster than the highly optimized GPU Thrust radix sort that directly manipulates the binary representation of keys. Our implementation is robust to different distributions and entropy levels of keys and scales almost linearly with the input size. These results indicate that multi-way techniques in general and sample sort in particular achieve substantially better performance than two-way merge sort and quicksort.

## 1 Introduction

Sorting is one of the most widely researched computational problems in computer science. It is an essential building block for numerous algorithms, whose performance depends on the efficiency of sorting. It is also an internal primitive utilized by database operations, and therefore, any application that uses a database may benefit from an efficient sorting algorithm. Geographic information systems, computational biology, and search engines are further fields that involve sorting. Hence, it is of utmost importance to provide efficient sort primitives for emerging architectures, which exploit architectural attributes, such as increased parallelism that were not available before.

The main trend in current CPU architecture is a continously increasing chip-level parallelism. Whereas quadcore processors have become already a commonplace, in the years to come, core numbers are likely to follow Moore's law. This trend to manycore processors is already realized in Graphical Processing Units (GPUs). Current NVidia GPUs, for example, feature up to $240$ scalar processing units per chip, which are directly programmable in C using the Compute Unified Device Architecture (CUDA)[8].

*Universität Karlsruhe (TH), Germany, `nikolaj.leischner@student.kit.edu`.
†Universität Karlsruhe (TH), Germany, `osipov@ira.uka.de`.
‡Universität Karlsruhe (TH), Germany, `sanders@ira.uka.de`.

In this paper we describe the design and implementation of a *sample sort* algorithm for such manycore processing units using CUDA, that in contrast to the approaches directly manipulating the binary representation of data (such as radix sort), requires a comparison function on keys only.

Our experimental study demonstrates that our sample sort is faster than all previously published comparison-based GPU sorting algorithms, and outperforms the state-of-the-art radix sort from Thrust library on $64$-bit integers and some nonuniform distributions of $32$-bit integers. It is robust to the commonly accepted set of distributions used for experimental evaluation of sorting algorithms [7] and peforms equally well for the whole range of input sizes.

One of the main reasons for a better performance of sample sort over quicksort and merge sort is that it needs less accesses to global memory since it processes the data in several multi-way phases rather than a larger number of two-way phases. Previous multi-way algorithms on GPUs, radix sorts from the CUDPP and Thrust libraries [12], directly manipulate the binary representation of keys, and thus, become inefficient when the keys are long or nonuniformly distributed.

Highly parallel GPU architectures support and in fact expect thousands of concurrent threads for optimal utilization of computational resources. Therefore, an efficient implementation has to expose this parallelism by carefully taking the architectural particularities of GPUs into account. To exploit coarse level parallelism, we impose a block-wise structure on the algorithm, that allows data-parallel processing of individual input tiles by blocks of fine-grained concurrent threads. We reduce the number of global memory accesses by storing memory intensive data structures in fast per-block memory. To expose fine-grained data-level parallelism, we adopt a technique for avoiding branch mispredictions previously used on commodity CPUs [11]. Our implementation is capable of processing large amounts of data by assigning a variable number of elements per thread. We balance out the computational load and memory latency by choosing the optimal multi-way distribution degree and the proper input size for a fallback to an alternative sorting method.

Although we implemented the algorithm using CUDA for GPUs, our techniques should be adaptable to other manycore architectures, in particular to systems that support OpenCL [1], which have a programming model similar to CUDA.

# 2 Parallel Computations using CUDA

To justify the particular design decisions that we made in our implementation, it is necessary to review specific architectural features of current NVidia GPUs reflected in the CUDA programming model together with general performance guidelines and design patterns.

Initially, Graphics Processing Units implemented a fixed-function graphic pipeline and evolved into a programmable parallel processor. Not only did NVidia's Tesla architecture, introduced in 2006 in the GeForce 8800 GPU, enable applications written in C, but also gave access to some capabilities of the GPU, that were not yet available through graphics APIs. Most notably, it featured fast per-multiprocessor shared memory and scattered writing to GPU memory.

Current NVidia GPUs feature up to $30$ streaming multiprocessors (SMs) each of which containing $8$ SPs, i.e., up to $240$ physical cores. However, they require a minimum of around $5\,000$–$10\,000$ threads to fully utilize hardware and hide memory latency. A single SM has $2048$ $32$-bit registers, for a total of $64$KB of register space and $16$KB on-chip shared memory that has very low latency

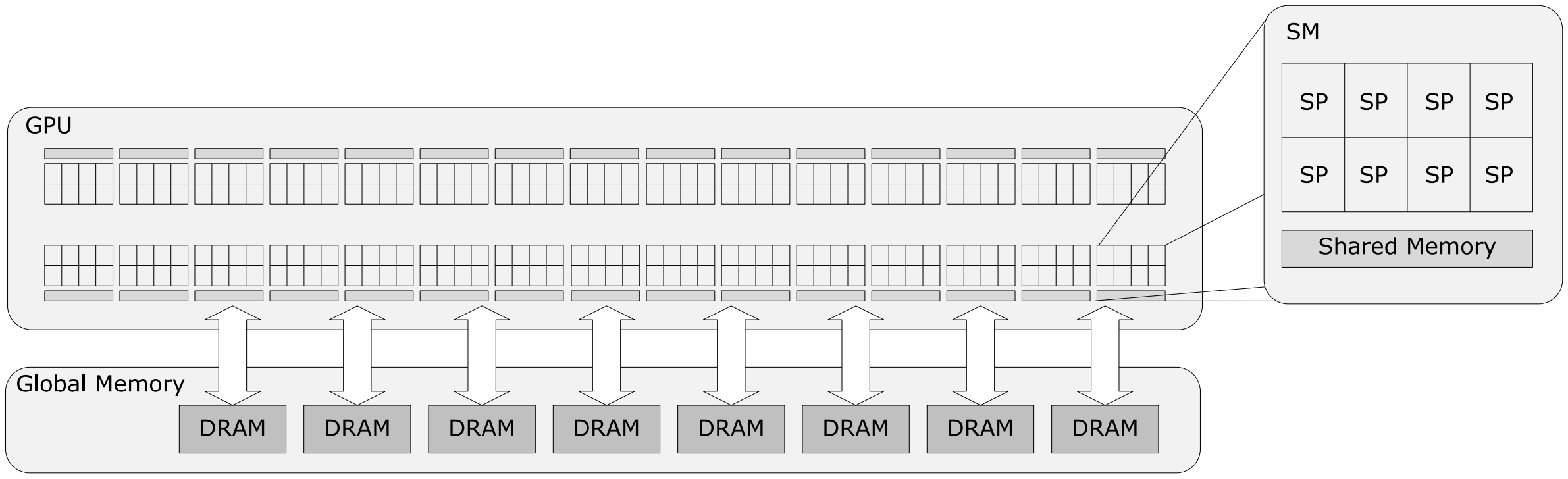

Figure 1: Tesla architecture

and high bandwidth similar to L1 cache.

To create, execute and manage hundreds of threads, the SM uses a processor architecture referred to as single-instruction multiple-thread (SIMT). An SM's SIMT instruction unit executes threads in *warps* – groups of $32$ threads. Each SM in a current NVidia GPUs manages a pool of $32$ warps, for a total of $1024$ threads. At each cycle, the scheduler selects a warp from the pool and executes an SIMT instruction on it involving almost no overhead for scheduling.

To achieve peak performance, an efficient algorithm should take certain SIMT attributes into careful consideration:

**conditional branching:** threads within a warp are executed in an SIMD fashion, i.e., if threads diverge on a conditional statement, both branches are executed one after another. Therefore, an SIMT processor realizes its full efficiency when all warp threads agree on the same execution path. Divergence between different warps, however, introduces no performance penalty;

**shared memory:** SIMT multiprocessors have on-chip memory (currently up to $16$KB) for low-latency access to data shared by its cooperating threads. Shared memory is orders of magnitude faster than the global device memory. Therefore, designing an algorithm that exploits fast memory is often essential for higher performance;

**coalesced global memory operations:** aligned load/store requests of individual threads of a warp to the same memory block are coalesced into fewer memory accesses than to separate ones. Hence, an algorithm that uses such access patterns is often capable of achieving higher memory throughput.

The CUDA programming model provides the means for a developer to map a computing problem to such a highly parallel processing architecture. A common design pattern is to decompose the problem into many data-independent sub-problems that can be solved by groups of cooperative parallel threads, referred to in CUDA as *thread blocks*. Such a two-level parallel decomposition maps naturally to the SIMT architecture: a block virtualizes an SM processor and concurrent threads within the block are scheduled for execution on the SPs of one SM.

A single CUDA computation is in fact similar to the SPMD (single-program multiple-data) software model: a scalar sequential program, a *kernel*, is executed by a set of concurrent threads, that constitute a grid of blocks. Overall, a CUDA application is a sequential CPU, *host*, program that launches kernels on a GPU, *device*, and specifies the number of blocks and threads per block for each kernel call.

# 3 Related work

Since sorting is one of the most widely studied areas in computer science, there is too much work done in the area to review it here. Therefore, we mainly focus on algorithms for parallel manycore architectures that are most relevant to our work.

Until recently, refined versions of quicksort were considered among the fastest general purpose sorting algorithms for single core machines used in practice [10]. However, the emergence of current generation CPUs featuring several cores, large caches and an SIMD instruction set, turned the focus on more cache efficient divide-and-conquer approaches that were able to expose a higher level of parallelism. Indeed, to our knowledge there is no efficient quicksort implementation, which exploits SIMD instructions. Moreover, despite having perfect spatial locality, quicksort requires at least $\log(n/M)$ scans until subproblems fit into a cache of size $M$.

A general divide-and-conquer technique can be described in three steps: the input is recursively split into $k$ tiles while the tile size exceeds a fixed size $M$, individual tiles are sorted independently and merged into the final sorted sequence. Most divide-and-conquer algorithms are based either on a $k$-way distribution or a $k$-way merge procedure. In the former case, the input is split into tiles that are delimited by $k$ ordered splitting elements. The sorted tiles form a sorted sequence, thus making the merge step superfluous. As for a $k$-way merge procedure, the input is evenly divided into $\log_k n/M$ tiles, that are sorted and $k$-way merged in the last step. In contrast to two-way quicksort or merge sort, multi-way approaches perform $\log_k n/M$ scans through the data (in expectation for $k$-way distribution).

This general pattern gives rise to several efficient manycore algorithms varying only in the way they implement individual steps. For instance, in a multicore gcc sort routine [14], each core gets an equal-sized part of the input (thus $k$ is equal to the number of cores), sorts it using introsort [10], and finally, cooperatively $k$-way merges the intermediate results.

Another recently published multicore algorithm following the same pattern additionally uses SIMD instructions [5]. For a CPU cache of size $M$ it divides the input into $n/M$ equal-sized parts, sorts them using bitonic sort and SIMD instructions in cache [2], and finally multi-way merges results. To our knowledge this algorithm is the fastest published multicore sorting approach at least for the key types reported in the paper.

Though GPUs are generally better suited for computational rather than combinatorial problems, new architectural capabilities of graphics processors brought considerable attention to sorting on GPUs.

For example, Harris et al. [13] developed an efficient *scan* (prefix sum) primitive – an essential building block for data parallel computation with numerous applications. By reducing counting sort to a number of scan primitives, Harris et al. [12] were able to design an efficient radix sort algorithm. They also gave the first implementation of quicksort that was based on a segmented scan primitive [13]. However, high overhead induced by this approach led to a sort that was not competitive to an explicit partitioning scheme, that was used in an alternative implementation by Cederman and Tsigas [3].

One of the first GPU-based two-way merge sort algorithms appeared as the second phase of a two step approach by Sintorn and Assarsson [15]. The algorithm divides the input into $n/4$ tiles, sorts all of them and merges the chunks in $\log(n/4)$ iterations by assigning one thread to each pair of sorted sequences. To improve parallelism in the last iterations, it initially partitions the input into sufficiently many tiles assuming that the keys are uniformly distributed. Another recent approach is bbsort [4] based on initial partitioning similar to that of hybrid sort.

As for comparison-based sorting algorithms on GPUs in general, the fastest algorithm described in the literature currently is a two-way merge sort by Harris et al. [12]. It divides the input into $n/256$ tiles, sorts them using odd-even merge sort [2] and two-way merges the results in $\log(n/256)$ iterations. In contrast to hybrid sort several threads can work cooperatively on merging two sequences, therefore eliminating the need for preliminary partitioning.

In the same work [12] Harris et al. presented a very efficient variant of radix sort, which is superior to all other GPU and CPU sorting algorithms at least for $32$-bit integer keys and key-value pairs.

# 4 Algorithm Overview

```
SampleSort(e = ⟨e_1, ..., e_n⟩, k)
begin
    if n < M then return SmallSort(e)
    choose a random sample S = S_1, ..., S_{ak-1} of e
    Sort(S)
    ⟨s_0, s_1, ..., s_k⟩ = ⟨-∞, S_a, ..., S_{a(k-1)}, ∞⟩
    for 1 ≤ i ≤ n do
        find j ∈ {1, ..., k}, such that s_{j-1} ≤ e_i ≤ s_j
        place e_i in bucket b_j
        return Concatenate(SampleSort(b_1, k), ..., SampleSort(b_k, k))
    end
end
```

**Algorithm 1**: Serial sample sort

Sample sort is considered to be one of the most efficient comparison-based algorithms for distributed memory architectures. Its sequential version is probably best described in pseudocode, see Algorithm 1. The oversampling factor $a$ trades off the overhead for sorting the splitters and the accuracy of partitioning.

The splitters partition input elements into $k$ buckets delimited by successive splitters. Each bucket can then be sorted recursively and their concatenation forms the sorted output. If $M$ is the size of the input when `SmallSort` is applied, the algorithm requires $\mathcal{O}(\log_k n/M)$ $k$-way distribution phases in expectation until the whole input is split into $n/M$ buckets. Using quicksort as a small sorter leads to an expected execution time of $\mathcal{O}(n \log n)$.

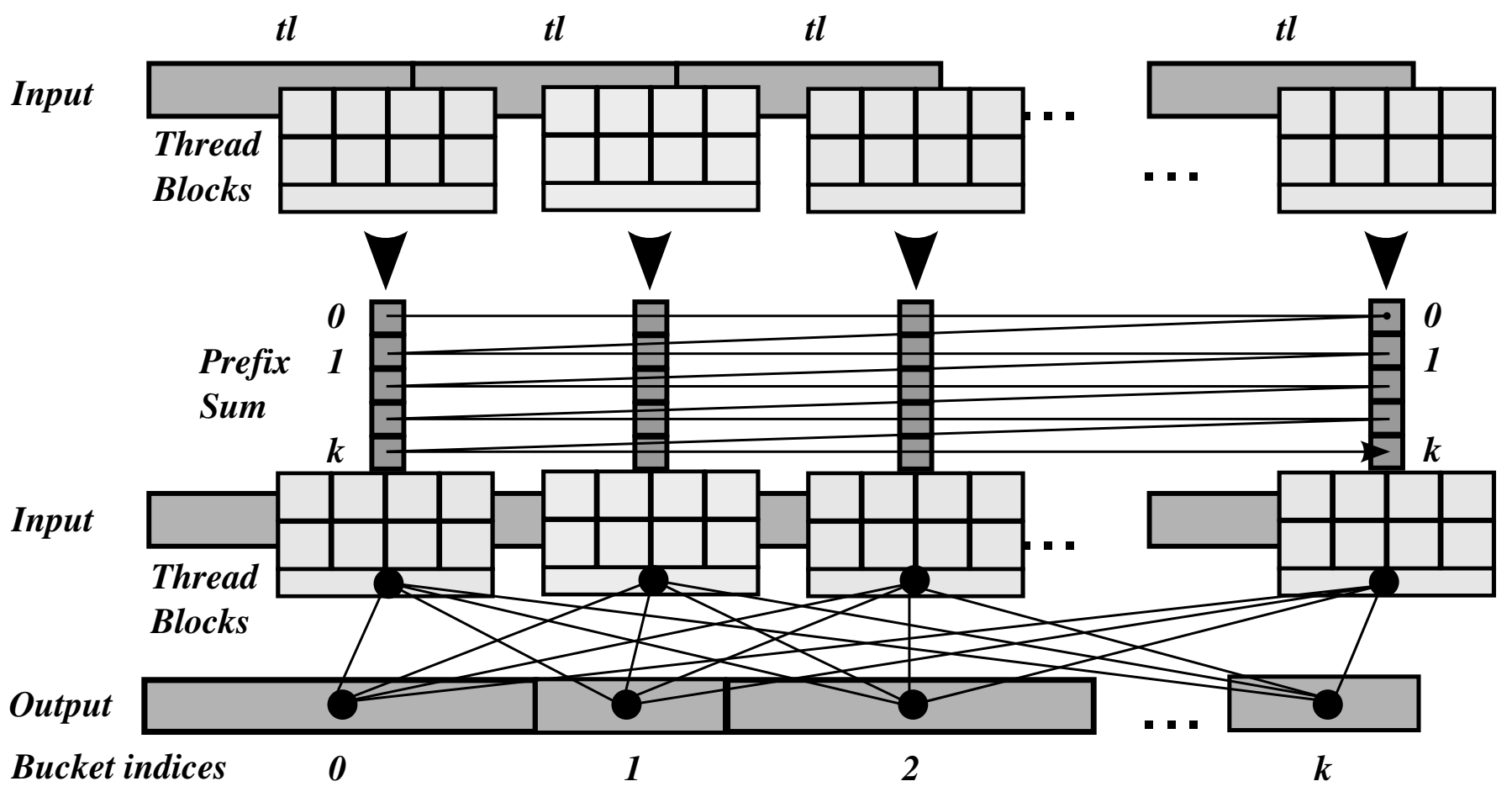

Figure 2: An iteration of $k$-way distribution

Haris et al. [12] pointed out, that their main reason for favoring merge sort over sample sort was its implicit load balancing. They considered it more beneficial than sample sort's avoidance of interprocessor communication. Since the bucket sizes heavily depend on the quality of the splitters, this argument is certainly true. However, sufficiently large random samples yield provably good splitters independent of the input distribution. Therefore, one should not overestimate the impact of load balancing on the performance.

On the other hand, due to the high cost of global memory accesses on GPUs, multi-way approaches are more promising than two-way: Each $k$-way distribution phase requires only $\mathcal{O}(n)$ memory accesses. Expected $\log_k(n/M)$ passes are needed until buckets fit into fast GPU shared memory $M$. Thus, we can expect $\mathcal{O}(n \log_k (n/M))$ global memory accesses instead of $\mathcal{O}(n \log (n/M))$ required by two-way merge sort [12]. This asymptotical behavior motivated our work in the direction of $k$-way sorting algorithms, and sample sort in particular.

Before we describe the design of GPU sample sort we should mention that hybrid sort [15] and bbsort [4] involve a distribution phase assuming that the keys are uniformly distributed. The uniformity assumption simplifies partitioning, but makes these approaches not competitive to sample sort on nonuniform distributions, see Section 6. Moreover, though such distribution approaches are suitable for numerical keys, they are not comparison-based.

As explained in Section 2, in order to efficiently map a computational problem to a GPU architecture we need to decompose it into data-independent subproblems that can be processed in parallel by blocks of concurrent threads. Therefore, we divide the input into $p = \lceil n/(t \cdot \ell) \rceil$ tiles of $t \cdot \ell$ elements and assign one block of $t$ threads to each tile, thus each thread processes $\ell$ elements sequentially. Even though one thread per element would be a natural choice, such independent serial work allows a better balance of the computational load and memory latency, see Section 5.

A high-level design of a sample-sort's distribution phase, when the bucket size exceeds a fixed size $M$, can be described in $4$ phases corresponding to individual GPU kernel launches, see Figure 2.

**Phase 1.** Choose splitters as in Algorithm 1.

**Phase 2.** Each thread block computes the bucket indices for all elements in its tile, counts the number of elements in each bucket and stores this per-block $k$-entry histogram in global memory.

**Phase 3.** Perform a prefix sum over the $k \times p$ histogram tables stored in a column-major order to compute global bucket offsets in the output, for instance the Thrust implementation [13].

**Phase 4.** Each thread block again computes the bucket indices for all elements in its tile, computes their local offsets in the buckets and finally stores elements at their proper output positions using the global offsets computed in the previous step.

At first glance it seems to be inefficient to do the same work in phases $2$ and $4$. However, we found out that storing the bucket indices in global memory (as in [11]) was not faster than just recomputing them, i.e., the computation is memory bandwidth bounded so that the added overhead of $n$ global memory accesses undoes the savings in computation.

We also tried more involved strategies like sorting the elements in shared memory by their bucket index and storing the sorted sequence in global memory in Phase $2$ (as in [12]) in order to avoid recomputing bucket indices and to improve coalesced scattered writing (see Section 2) in Phase 4. However, it was still slower than the simple approach we present here. While unstructured memory accesses considerably decrease the effective memory bandwidth, the latency can be hidden by interleaving memory accesses and computation.

For buckets of size less than $M$, one can use any GPU sorting algorithm. In our final implementation we chose to use an adaptation of quicksort by Cederman and Tsigas [3].

In the following section, we give a detailed description of each phase of the algorithm including design choices we made motivated by architectural attributes and performance guidelines reviewed in Section 2.

# 5 Implementation Details

**Phase 1.** We take a random sample $S$ of $a \cdot k$ input elements using a simple GPU LCG random number generator that takes its seed from the CPU Mersenne Twister [9]. Then we sort it, and place each $a$-th element of $S$ in the array of splitters $bt$ such that they form a complete binary search tree with $bt[1] = s_{k/2}$ as the root. The left child of $b[j]$ is placed at the position $2j$ and the right one at $2j + 1$, see Algorithm 2.

**Phase 2.** To speed up the traversal of the search tree and save accesses to global memory, each block loads $bt$ into its fast private memory shared by its threads (see Section 2).

To find a bucket index for an element we adopt a technique that originally was used to prevent branch mispredictions impeding instruction-level parallelism on commodity CPUs [11]. In our case, it allows avoiding conditional branching of threads while traversing the search tree, refer to Algorithm 2. Indeed, a conditional increment in the loop is replaced by a predicated instruction. Therefore, threads concurrently traversing the search tree do not diverge, thus avoiding serialization, see details on conditional branching in Section 2. Since $k$ is known at compile time, the compiler can unroll the loop, which further improves the performance.

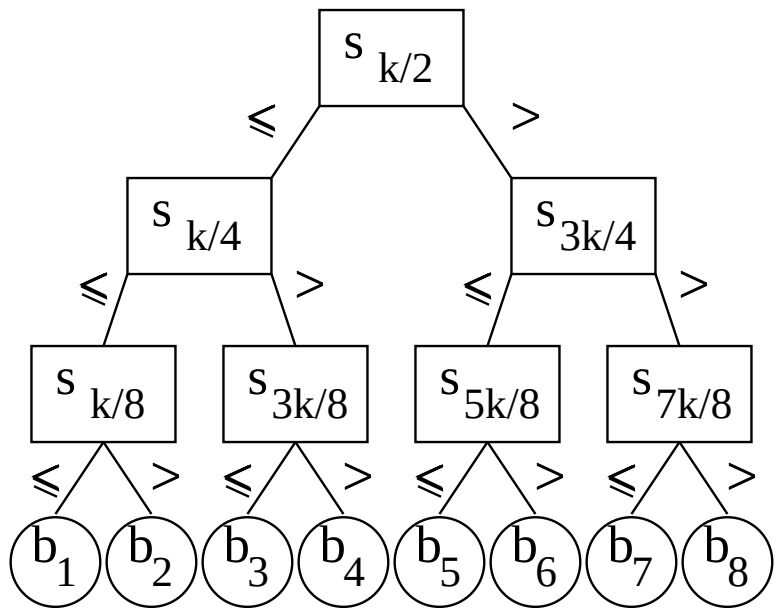

$$bt = \langle s_{k/2}, s_{k/4}, s_{3k/4}, s_{k/8}, s_{3k/8}, s_{5k/8}, s_{7k/8} \ldots \rangle$$

`TraverseTree(`$e_i$`)`
**begin**
  $j := 1$
  **repeat** $\log k$ times $j := 2j + (e_i > bt[j])$ `// left or right child?`
  $j := j - k + 1$ `// bucket index`
**end**

**Algorithm 2**: Serial search tree traversal

Having computed the bucket indices for its elements, each thread block counts the sizes of the resulting buckets by incrementing counters in its shared memory. To avoid the contention of concurrent threads we exploit atomic add instructions. Since many threads may increment the same counter, we improve parallelism by splitting threads into groups and use individual counter arrays per group. We found $8$ arrays to be a good compromise between overhead for handling several arrays and a lack of parallelism when only one array is used. On hardware that does not support atomic operations we can explicitly avoid conflicts by using a single thread per group for counting.

Per-block $k$-entry histograms in global memory resulting from the vector sum computation on the bucket size arrays form the output of this phase.

Note that the larger the tiles are, i.e., the larger the parameters $t$ and $\ell$ are, the less output is produced by this phase, reducing the number of global memory accesses. On the other hand side, large tiles decrease parallelism. Therefore the parameter $\ell$ allows a flexible trade-off between these two effects.

**Sorting buckets.** We delay the sorting of buckets until the whole input is partitioned into buckets of size at most $M$. Since the number of buckets grows with the input size, it is larger than the number of processors in most of the cases. Therefore, we can use a single thread block per bucket without sacrificing exploitable parallelism. To improve load-balancing we schedule buckets for sorting ordered by size.

In order to sort buckets efficiently we split them into chunks that fit into shared memory, which then can be sorted without expensive accesses to the global memory. Initially we employed a sample sort based algorithm for this purpose. But we found that it performed slightly worse than our optimized quicksort implementation based on the GPU quicksort by Cederman and Tsigas [3], which we therefore use in our final implementation. Quicksort does not cause any serialization of work, except for pivot selection and stack operations. Additionally, its consumption of registers and shared memory is modest.

For sequences that fit into shared memory, we use an odd even merge sorting network [2]. In our experiments we found it to be faster than the bitonic sorting network and other approaches like a parallel merge sort.

**Parameters.** The choice of parameters balances out such attributes as scalability, exposed parallelism and memory latency. By setting $k = 128$ and $M = 2^{17}$ we trade-off the nonuniformity of bucket sizes produced by the $k$-way distribution against the better performance of quicksort on small instances. This way we achieve an almost uniform sorting rate throughout the whole input size range.

The oversampling factor $a = 30$ for smaller keys and $a = 15$ for $64$-bit integers produces a good quality sample and still allows sorting in the second phase to be completely performed in shared memory, thus inducing almost no overhead when compared to the smaller factors.

By choosing $t = 256$ threads per block and $\ell = 8$ elements per thread, we achieve a compromise between the parallelism exposed by the algorithm, the amount of data $(n \cdot k)/(t \cdot \ell)$ written in the second phase and memory latency in the fourth phase.

# 6 Experimental Study

We report experimental results of our sample sort implementation on sequences of floats, $32$-bit and $64$-bit integers and key-value pairs where both keys and values are $32$-bit integers. We compare the performance of our algorithm to a number of existing GPU implementations including: state-of-the-art Thrust and CUDPP radix sorts and Thrust merge sort [12], as well as quicksort [3], hybrid sort [15] and bbsort [4]. Since most of the algorithms do not accept arbitrary key types, we omit them for the inputs they were not implemented for. We have not included approaches based on graphics APIs in our benchmark, bitonic sort in particular [6], since they are not competitive to the CUDA-based implementations listed above.

Our experimental platform is an Intel Q6600 $2.4$ GHz quad-core machine with $8$GB of memory. We used an NVidia Tesla C1060 that has $30$ Multiprocessors, each containing $8$ scalar processors, for a total of up to $240$ cores on chip. In comparison to commodity NVidia cards, the Tesla C1060 has a larger memory of $4$GB, that allows a better scalability evaluation. We compiled all implementations using CUDA $2.3$ and gcc $4.3.2$ on $64$-bit Suse Linux $11.1$ with optimization level -O3.

We do not include the time for transferring the data from host CPU memory to GPU memory, since sorting is often used as a subroutine for large-scale GPU computations.

For the performance analysis we used a commonly accepted set of distributions motivated and described in [7].

**Uniform.** A uniformly distributed random input in the range $[0, 2^{32} - 1]$.

**Gaussian.** A gaussian distributed random input approximated by setting each value to an average of $4$ random values.

**Bucket Sorted.** For $p \in \mathbb{N}$, the input of size $n$ is split into $p$ blocks, such that the first $n/p^2$ elements in each of them are random numbers in $[0, 2^{31}/p - 1]$, the second $n/p^2$ elements in $[2^{31}/p, 2^{32}/p - 1]$, and so forth.

**Staggered.** For $p \in \mathbb{N}$, the input of size $n$ is split into $p$ blocks such that if the block index is $i \leq p/2$ all its $n/p$ elements are set to a random number in $[(2i - 1)2^{31}/p, (2i)(2^{31}/p - 1)]$.

**Deterministic Duplicates.** For $p \in \mathbb{N}$, the input of size $n$ is split into $p$ blocks, such that the elements of the first $p/2$ blocks are set to $\log n$, the elements of the second $p/4$ processors are set to $\log(n/2)$, and so forth.

We use $p = 240$ (the number of scalar processors of a Tesla C1060) and the Mersenne Twister [9] as a source of uniform random values.

**Key-value pairs.** Since the best comparison-based sorting algorithm, Thrust merge sort, is designed for key-value pairs only, we can fairly compare it to our sample sort only on this input type. On uniformly distributed keys, our sample sort implementation is at least $25\%$ faster, and achieves on average a $68\%$ higher performance than Thrust merge sort. We do not depict all distributions on key-value pairs, but rather mention the worst case behavior of our implementation on sorted sequences. Sample sort is at least as fast as Thrust merge sort, and still is $30\%$ better on average, see Figure 3.

Similarly to radix sort on commodity CPUs, CUDPP radix sort is considerably faster than the comparison-based sample and merge sort on $32$-bit integer keys. However, on low level entropy inputs, such as Deterministic Duplicates, see Figure 3, even for such low length key types, radix sort is outperformed by sample sort.

**64-bit integer keys.** With the growth of the key length, radix sort's dependence on the binary key representation makes Thrust radix sort (the only implementation accepting $64$-bit keys) not competitive to sample sort. On uniformly distributed keys, our sample sort is at least $63\%$ and on average $2$ times faster than Thrust radix. On a sorted sequence, which is the input when our implementation performs worst, its sorting rate does not deviate significantly from the uniform case, see Figure 4

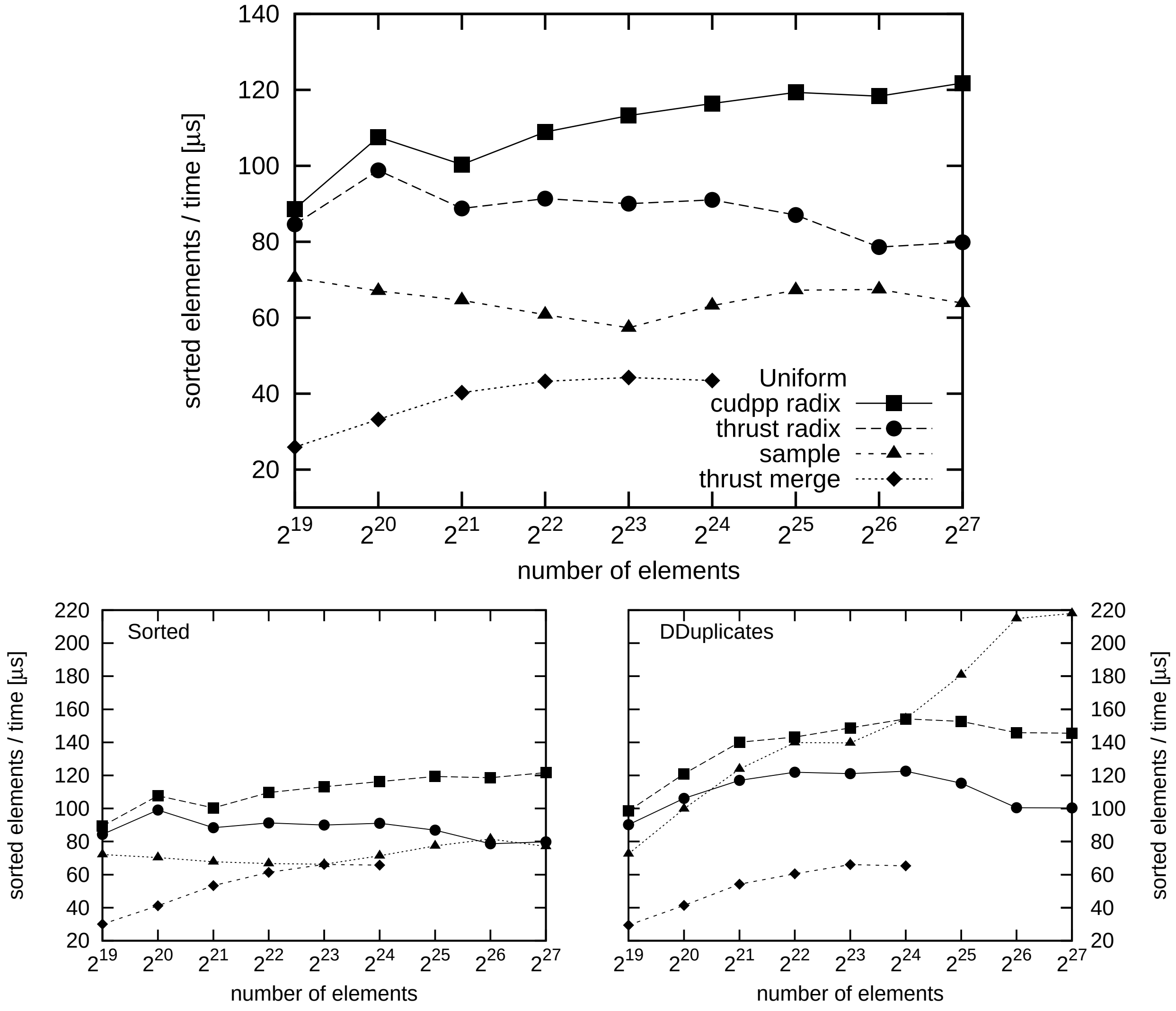

Figure 3: Sorting rates on key-value pairs.

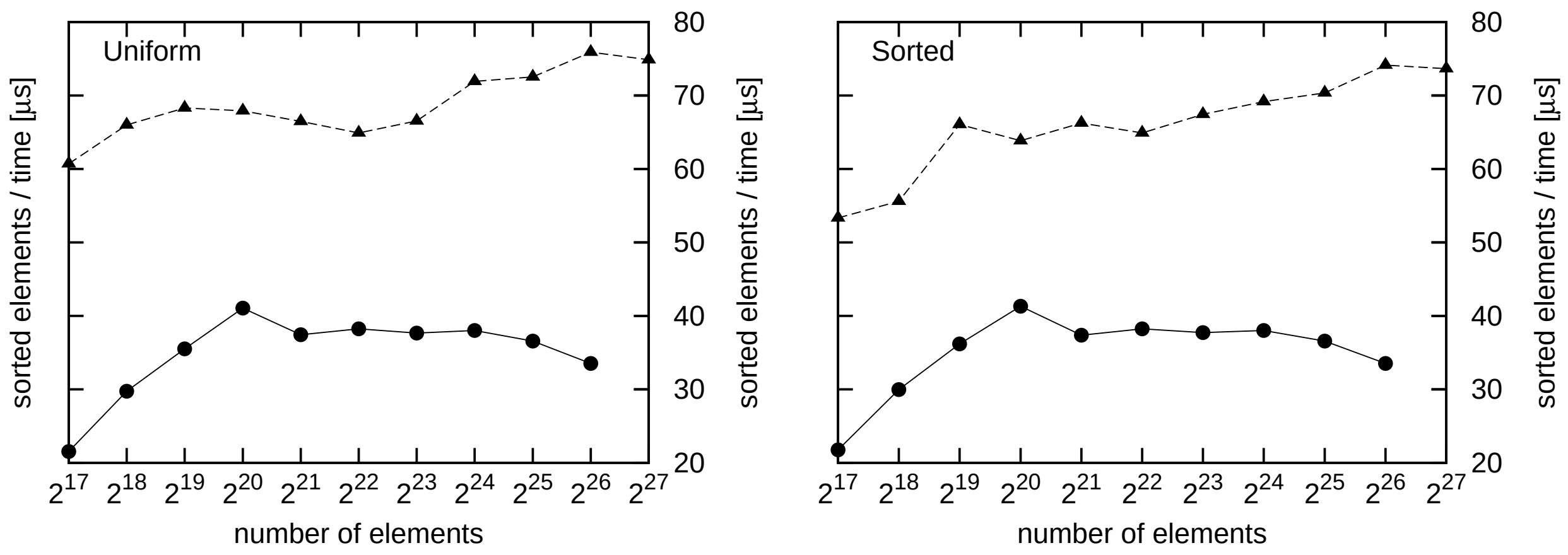

Figure 4: Sorting rates on 64-bit integer keys.

**32-bit integer keys.** Since the majority of GPU sorting implementations are able to sort $32$-bit integers we report sample sort's behavior on all distributions listed above, refer to Figure 5. We include hybrid sort results on floats, since it is the only key type accepted by this implementation, and the sorting rates of other algorithms on floats are similar to the ones on integer inputs.

Low length key type allows both implementations of radix sort to outperform all algorithms similar to the $32$-bit integer key-value pairs case. While sample sort demonstates the fastest and still robust performance over all other approaches, except for radix sorts. In particular, it is on average more than $2$ times faster than quicksort and hybrid sort for uniform distribution. Due to the uniformity assumption, and hence, a reduced computational cost involved as we mentioned in Section 4, bbsort is competitive, but still outperformed by our implementation. On the other hand side, the performance of bbsort as well as hybrid sort on Bucket and Staggered distributions significantly degrades when compared to the uniform case. Moreover, on the Deterministic Duplicates input, bbsort becomes completely innefficient, while hybrid sort crashes.

Sample sort is robust with respect to all tested distributions and performs almost equally well on all of them. It demonstrates a sorting rate close to constant, i.e., scales almost linearly with the input size. A higher level of parallelism, and hence, a better possibility of hiding memory latency on large inputs dominate the logarithmic factor in the runtime complexity. .

**Exploring bottlenecks.** Figure 6 reports sorting rates of CUDPP and Thrust radix sorts as well as Thrust merge sort and our sample sort on two different GPUs: NVidia Tesla C1060 and Zotac GTX 285. These GPUs have the same number of scalar processors, but the GTX 285 is clocked at $1.476$GHz and Tesla at $1.296$GHz, i.e., $13$% slower. The measured memory bandwidth of GTX285 is $124.7$GB/s, while Tesla's is $70$% slower, only $73.3$GB/s. The average improvements of CUDPP and Thrust radix sorts on the GTX285 are $30$% and $25$% respectively, while Thrust merge and sample sorts improve just by $18$%. This indicates that none of the algorithms is solely computationally or memory bandwidth bounded. However, the larger improvement for both radix sorts suggests that they are rather memory bandwidth bounded, while merge and sample sort are more computationally bounded.

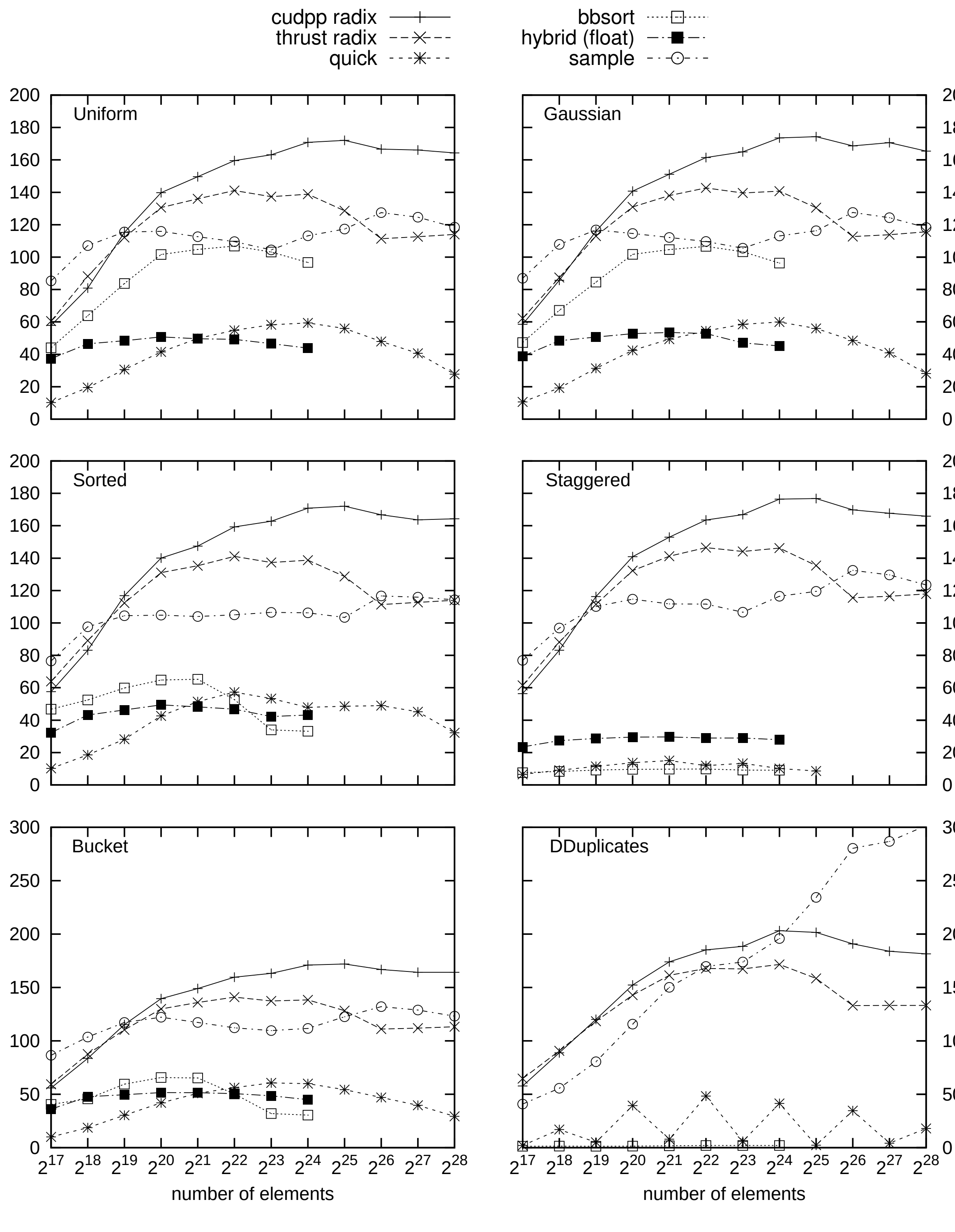

Figure 5: Sorting rates on 32-bit integers.

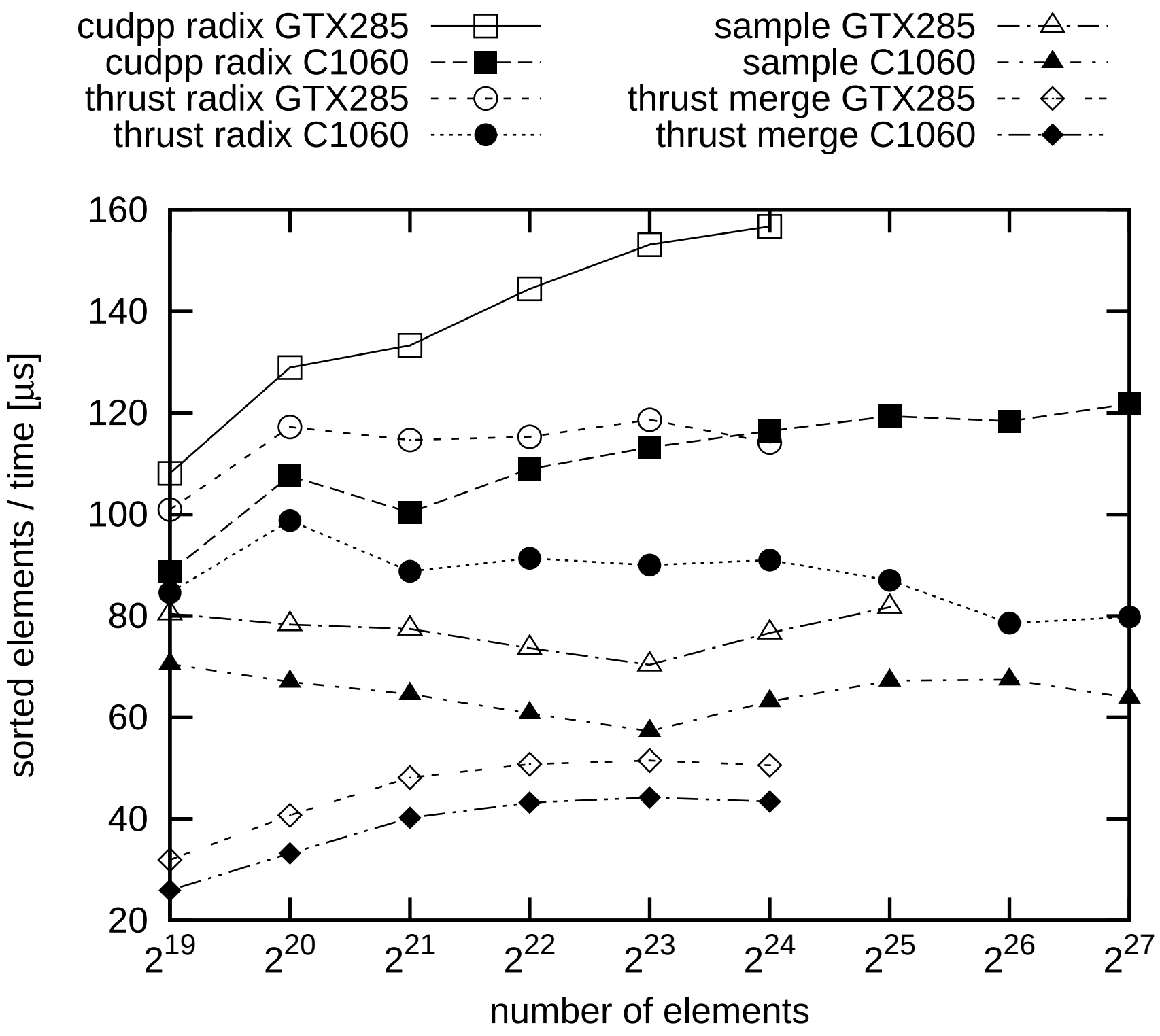

Figure 6: Sorting rates on uniform key-value pairs on Tesla C1060(black dots) and Zotac GTX285 (white dots)

# 7 Conclusion

In this paper we demonstrated the design and implementation of a sample sort algorithm that outperforms all previous comparison-based approaches as well as algorithms directly manipulating the binary representation of data on longer keys. Our experimental study demonstrates that our implementation is robust to different input types, thus disproving a previous conjecture that the performance of sample sort on GPUs is highly dependent on the distribution of keys [12].

This is an indication that multi-way algorithms in general and the sample sort's multi-way distribution in particular are superior on GPUs to two-way aproaches, which require more passes to process the data. In this respect we believe that the performance of other multi-way approaches, foremost multi-way merge sort, should be explored further.

Though we implemented the algorithm using CUDA for GPUs, our techniques should also suit other manycore architectures well.

# References

[1] Opencl. `http://www.khronos.org/opencl/`.

[2] K. E. Batcher. Sorting networks and their applications. In *Proc. AFIPS Spring Joint Computing Conference*, volume 32 of *AFIPS Conference Proceedings*, pages 307–314, 1968.

[3] D. Cederman and P. Tsigas. A practical quicksort algorithm for graphics processors. In *Proc. European Symposium on Algorithms (ESA)*, volume 5193 of *LNCS*, pages 246–258, 2008.

[4] S. Chen, J. Qin, Y. Xie, J. Zhao, and P.-A. Heng. A fast and flexible sorting algorithm with cuda. In *ICA3PP*, volume 5574 of *LNCS*, pages 281–290, 2009.

[5] J. Chhugani, A. D. Nguyen, V. W. Lee, W. Macy, M. Hagog, Y.-K. Chen, A. Baransi, S. Kumar, and P. Dubey. Efficient implementation of sorting on multi-core simd cpu architecture. *PVLDB*, 1(2):1313–1324, 2008.

[6] N. K. Govindaraju, J. Gray, R. Kumar, and D. Manocha. Gputerasort: high performance graphics co-processor sorting for large database management. In *Proc. ACM SIGMOD Int'l Conference on Management of Data*, pages 325–336, 2006.

[7] D. R. Helman, D. A. Bader, and J. JáJá. A randomized parallel sorting algorithm with an experimental study. *J. of Parallel and Distributed Computing*, 52(1):1–23, 1998.

[8] E. Lindholm, J. Nickolls, S. F. Oberman, and J. Montrym. Nvidia tesla: A unified graphics and computing architecture. *IEEE Micro*, 28(2):39–55, 2008.

[9] M. Matsumoto and T. Nishimura. Mersenne twister: A 623-dimensionally equidistributed uniform pseudo-random number generator. *ACM Transactions on Modeling and Computer Simulation*, 8(1):3–30, 1998.

[10] D. R. Musser. Introspective sorting and selection algorithms. *Software: Practice and Experience*, 27(8):983–993, 1997.

[11] P. Sanders and S. Winkel. Super scalar sample sort. In *Proc. European Symposium on Algorithms (ESA)*, volume 3221 of *LNCS*, pages 784–796. Springer, 2004.

[12] N. Satish, M. Harris, and M. Garland. Designing efficient sorting algorithms for manycore gpus. In *Proc. Int'l Symposium on Parallel and Distributed Processing (IPDPS)*, 2009.

[13] S. Sengupta, M. Harris, Y. Zhang, and J. D. Owens. Scan primitives for gpu computing. In *Proc. ACM SIGGRAPH/EUROGRAPHICS Conference on Graphics Hardware*, pages 97–106, 2007.

[14] J. Singler, P. Sanders, and F. Putze. Mcstl: The multi-core standard template library. In *Proc. Int'l Conference on Parallel Processing (Euro-Par)*, volume 4641 of *LNCS*, pages 682–694, 2007.

[15] E. Sintorn and U. Assarsson. Fast parallel gpu-sorting using a hybrid algorithm. *J. of Parallel and Distributed Computing*, 68(10):1381–1388, 2008.